# Quantitative constraints to the complete state of stress from the combined borehole and focal mechanism inversions: Fox Creek, Alberta

Luyi W. Shen[1,2*], Douglas R. Schmitt[2,3], Kristine Haug[1]

[1]Alberta Geological Survey, Alberta Energy Regulator

[2]Institute of Geophysical Research, Department of Physics, University of Alberta

[3]now at Department of Earth, Atmospheric, and Planetary Sciences, Purdue University

[*]Correspondence to luyi@ualberta.ca

*Abstract*

We develop a quantitative predictive model for the state of stress of a volume of crust encompassing the Duvernay Formation with a spatial extent of over 150 km × 150 km centred near the municipality of Fox Creek. An average direction of the maximum horizontal compression $S_H$ is of N43°E, determined from analysis of 20 borehole image logs. A model of the vertical stress $S_V$, corrected for topographic variations, is provided from the integration of a 3D density volume constructed from 1125 density logs. The minimum horizontal compressions $S_h$ and pore pressures $P_P$ are evaluated through analysis of 57 well tests carried out within the Duvernay Formation; 3D models developed by kriging of the observed values. Stress inversion of the focal mechanism solutions for the earthquakes nearby validated the assumed Andersonian stress regime and provided the shape-ratio of the stress which further allowed estimation of maximum horizontal compression $S_H$. A program is provided allowing the user to calculate the full set of components necessary to describe the states of in-situ stress for a location of interest near the Duvernay formation within our study area.

*1* **Introduction**

Quantitative knowledge of the state of in-situ stress at many scales within the Earth's crust is key to understanding plate motions, underground construction safety, borehole integrity and, fault stability. The need to obtain such information has intensified by societal concerns associated with anthropogenically induced seismicity related to hydraulic fracturing in the development of unconventional resources. Slip-tendency of a fault is fundamentally governed by the traction force on the faulting planes and fault's constitutive behaviors. The central paradigm for such stability analyses assumes that displacement along a planar surface in the Earth occurs if the combination of resolved stresses and fluid pressure overcomes a Mohr-Coulomb frictional failure criterion (Hubbert and Rubey, 1959). This requires that the total normal $\sigma_N$ and shear $\tau$ tractions resolved onto this plane and the pore pressure $P_P$ are known with slip occurring once (Célérier et al., 2012)

$$\tau \geq \mu(\sigma_N - P_P) + S_o \qquad (1)$$

where $\mu$ is the coefficient of friction and $S_o$ is the cohesion of the plane; the compression positive sign convention is used. Calculation of $\tau$ and $\sigma_N$ on any arbitrarily oriented plane (Jaeger et al., 2007) requires the knowledge of the stress tensor $\sigma$.

Usually, the key limiting factor in carrying out stability analyses is the lack of quantitative information on the stress state. In this work, we aim to develop a complete quantitative Andersonian (1951) stress state model with consisting of the magnitudes of the vertical $S_V$, least horizontal $S_h$ and greatest horizontal $S_H$ principal stresses, and the $S_H$ orientation $\phi$ in an area that has recently experienced earthquakes induced by hydraulic fracturing operations.

$S_V$ is commonly assumed to be given by integration of the overlying gravitational load (McGarr and Gay, 1978). In certain areas, the stress orientation $\phi$ of the greatest horizontal compression $S_H$ may be inferred from existing compilations such as the World Stress Map (Heidbach et al., 2010). Finding the magnitudes of both $S_H$ and the least horizontal compression $S_h$ is difficult, and usually researchers make various assumptions based on extrapolations of quantitative results (Schwab et al., 2017; Weides et al., 2014), consistency with local earthquakes (Chang et al., 2010; Morris et al., 2017), frictional limits (Çiftçi, 2013; Schwab et al., 2017), constraints from various well tests (Chang et al., 2010; Konstantinovskaya et al., 2012), extrapolated empirical relationships (Adewole and Healy, 2017; Williams et al., 2016), or borehole stabilities (Peška and Zoback, 1995; Williams et al., 2016; Valley and Evans, 2019). As most of these authors indicate, the lack of proper measurements of the horizontal stress magnitudes, particularly with regards to $S_H$ for which direct measurement remains elusive, is the most significant source of uncertainties in their stability analyses.

Most stress compilations, too, have focused on tectonic stress directions over large areas. Regional studies that focus not only on stress directions but include quantitative determination of stress magnitudes from a representative number of boreholes remain rare (Bailey et al., 2016; Konstantinovskaya et al., 2012; Nelson et al., 2007; Streit and Hillis, 2004). Here, we focus on the development of a quantitative model for the magnitudes and directions of the stress tensor in an area of 150 km × 150 km centred at the municipality of Fox Creek, Alberta (Fig. 1a and b). The need to understand better the conditions leading to earthquakes that appear to have been produced by hydraulic fracturing of the unconventional Duvernay Formation in NW Alberta motivated our effort to develop a model allowing quantitative assessment of the states of in-situ stress. We extract magnitudes of $S_V$, $S_h$, and $P_p$ from geophysical well logs and transient pressure tests. Estimates of $\phi$ are obtained by interpretation of borehole deformation features and supported by stress inversion performed with reported focal mechanism solutions of nearby earthquakes. Subsequently, we constrain the range of $S_H$ by combining the shape-ratio $R$ determined from the stress inversion with the previously obtained $S_V$ and $S_h$. The dense measurements utilized in this sudy allows the development of a model which can be used to predict the quantitative stress values and $P_p$ in a crust volume encompassing the Duvernay formation and the epicenters of induced earthquakes.

## 1.1 Study Area

The study area from 53.5° to 55° N latitude and -118.5° to -115.5° W longitude (see Fig. 1a) includes the epicentres of felt earthquakes (see Fig. 1b) induced by hydraulic fracturing operations near the municipality of Fox Creek, Alberta (Schultz et al., 2018; Schultz et al., 2015; Schultz et al., 2017; Wang et al., 2016; Wang et al., 2017).

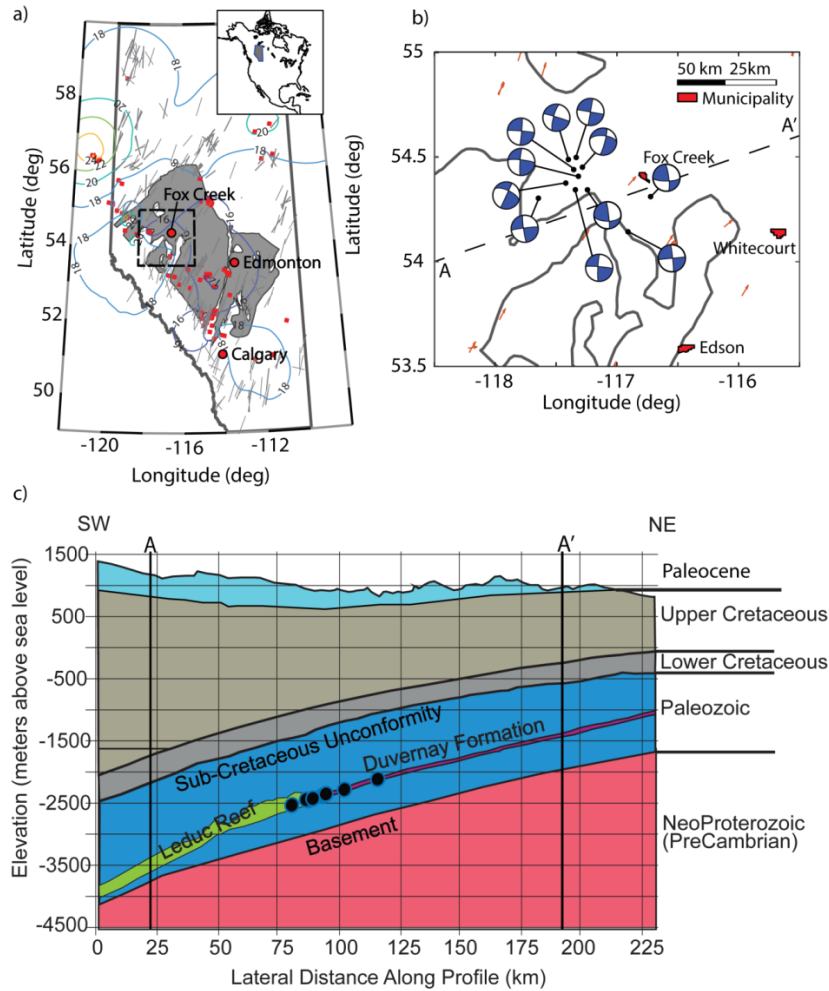

Fig. 1 a) Map of Alberta with lines giving $S_H$ directions in the WSM, red dots showing locations of $S_h$ magnitude measurements, and contours representing the estimated $S_h$-to-depth ratio as compiled by Haug and Bell (2016). The gray zone shows the areal extent of the Duvernay Formation. The black dashed lines enclose the area investigated in this study. b) An expanded view of the study area including epicentres and focal mechanism solutions for the felt earthquakes from 2013 to 2017 compiled by Schultz et al. (2017). The orange arrows indicate the $S_H$ orientations in the study area recorded by WSM. c) The cross-section of the stratigraphy (Branscombe et al., 2018) in the study area from A to A' as shown in b). The black dots along the Duvernay Formation represent the projection of earthquakes shown in b) assuming the earthquakes happen within the Duvernay Formation (Eaton et al., 2018; Schultz et al., 2017)

*1.2  Geological Structure*

The study area contains distinct packages of Neoproterozoic-Paleozoic and Phanerozoic sediment deposited, respectively, in a passive margin and foreland basin settings. These sediments overlie the stable North American cratonic basement constructed of Proterozoic crystalline rock (Fig. 1c). The Paleozoic portion of the sediments consists mostly of evaporites, carbonates, and shales deposited in open-marine environments, also including Frasnian age Leduc reef complexes contemporaneous with the Duvernay Formation that is a prominent target of hydraulic fracturing operations (Lyster et al., 2017). Development of the Alberta Foreland Basin commenced in the Late Jurassic-Early Cretaceous from the successive loading of overthrust sheets (Cant and Stockmal, 1989) with the basin filling with siliciclastics. In the study area, these siliciclastics were deposited onto the Paleozoic unconformity surface referred to as the Mississippian, or Sub Cretaceous Unconformity in the study region, and upwards of 2 km of these sediments may have since been removed by erosion (Wu, 1991). The craton and Paleozoic sediments display significant inclination; the top of the Duvernay Formation, for example, drops from ~ -1400 m below surface level (mbsl) in the NE to < -3000 mbsl in the SW of the study area (see Fig. 1c) while the top of the craton elevation declines from ~ -1800 mbsl to ~ -4000 mbsl (see Fig. 1c). In contrast, the surface elevation increases from ~750 mbsl to >2000 mbsl as it includes portions of the disturbed belt east of the Rocky Mountains.

The Paleozoic sediments burial subjected the organic-rich shale to conditions favourable for the formation of light hydrocarbon, most of which migrated eastward sourcing the many conventional carbonates and clastic hydrocarbon fields within the Western Canada Sedimentary Basin. Economic quantities of light hydrocarbons remain locked within the low permeability rock matrix of the organic-rich formations, and this has encouraged hydraulic fracturing activities in the Duvernay Formation (Preston et al., 2016), some of which are associated with induced seismicity.

Some faults have been seen in reflection seismic profiles more regionally outside the study area (Richards et al., 1994, Weides et al., 2014) or inferred from the structural mapping of geological tops (Green and Mountjoy, 2005). Analysis of textural seismic attributes of 3D seismic volumes at undisclosed locations in the region has hinted at the existence of subsurface lineaments that are possibly interpreted as faults (Chopra et al., 2017; Eaton et al., 2018) but to our knowledge most of these lineaments, many of which are not optimally oriented for slipping, have not been linked to any of the felt induced earthquakes.

*1.3    Fox Creek Induced Seismicity*

From December 2013 to 2016 more than 250 earthquakes with magnitudes $M_W > 2.5$ have been temporally correlated with hydraulic fracturing operations within Fox Creek Duvernay Formation (Schultz et al., 2017). All of these have occurred from ~15% of the wells within a relatively confined area of the administrative Kaybob assessment area (Schultz et al., 2018). It is also worth noting hydraulic fracturing operations within the Duvernay Formation about 200 km to the south (Willesden Green Field) have not induced any felt seismicity. Detailed reasons for this are as yet unknown, but spatial associations to possible basement faults associated with well mapped Devonian reef complexes (Schultz et al., 2016) or high pore pressures (Eaton, 2017; Eaton and Schultz, 2018; Fox and Soltanzadeh, 2015) were implicated. Statistical analyses further suggest that the volume of fluid injected during a given stimulation is a contributing factor, particularly when it is noted that no induced seismicity by hydraulic fracture operations before the first event in 2013 was detected (Schultz et al., 2018).

Schultz et al. (2017) provide a recent summary of the induced seismicity from 2013 to 2016 detailing about 250 events within 17 distinct clusters. The event's focal mechanism conjugate planes strike N-S/E-W predominantly and indicate strike-slip displacement with P-axis vectors pointing at an azimuth of 45° ± 5° consistent with expected regional horizontal stress directions (Bell and Grasby, 2012; Reiter et al., 2014) with the source mechanisms being largely double-couple. The earthquake epicentres are located with an accuracy of < 1 km, but the hypocenter depths remain poorly constrained at ~5±2 km (Schultz et al., 2017).

*1.4* **Prior Knowledge of Stress State**

The first efforts at mapping stress directions near the study area occurred in deep boreholes drilled in western Alberta. The analysis of borehole breakouts (BOs), using oriented calliper logging tools (dipmeters) by Bell and Gough (1979). Bell et al. (1994) gathered new BO stress direction measurements supplemented with a few quantitative estimates of $S_h$ magnitudes from a variety of well tests. This work has been periodically updated (Bell and Bachu, 2003; Bell and Grasby, 2012); with the final raw compilation recently available (Haug and Bell, 2016). Additional stress directions were also found within the literature, too, expanded information on stress orientations (Reiter et al., 2014) that mainly indicate NE-SW compression across Alberta; this direction is not necessarily always perpendicular to the disturbed belt of the Canadian Rocky Mountains as is often presumed.

Prior to the present study, there were 64 reported stress directions from Reiter et al. (2014) and included in the latest version of Word Stress Map (WSM, Heidbach et al., 2016) nearby (see supplementary material); 19 of these measurements are directly within the study area (see Fig. 1a). A small number of fracture closure (presumed to represent $S_h$) and $P_P$ (Woodland and Bell, 1989) are reported near the study area (see supplementary material), these data were incorporated into Haug and Bell's (2016) compilation which spatially covers the entire Alberta Basin. Sections of the Duvernay Formation appear to be highly over-pressured (Fox and Soltanzadeh, 2015). Shen et al. (2018a) attempted constraining the $S_H$ through the recorded width of borehole breakouts interpreted from image logs. However, this attempt is subject to some drawbacks including lack of knowledge on the rock mechanical parameters and $P_P$ at the rocks near the wellbore. Most of the borehole breakouts utilized in Shen et al. (2018a) are recorded in the segments above the Duvernay Formation. As such the full set of components of the in-situ stress, which are necessary for comprehensive characterization of the states of in-situ stress within our area of interests near Fox Creek, is still lacking. This lack of information motivates our study to constrain, as quantitatively as possible, the full stress tensor in the vicinity of the induced earthquakes.

## 2 Data and Methods

This study focuses on developing models for the directions and magnitudes that allow for extrapolation of the stress tensor to any points of interest within the study area. $P_P$, and $S_h$ magnitudes interpreted from a variety of borehole pressure tests (e.g., diagnostic fracture injection test (DFIT), static gradient survey, and flow/build up test) and values of $\phi$ extracted from image logs.

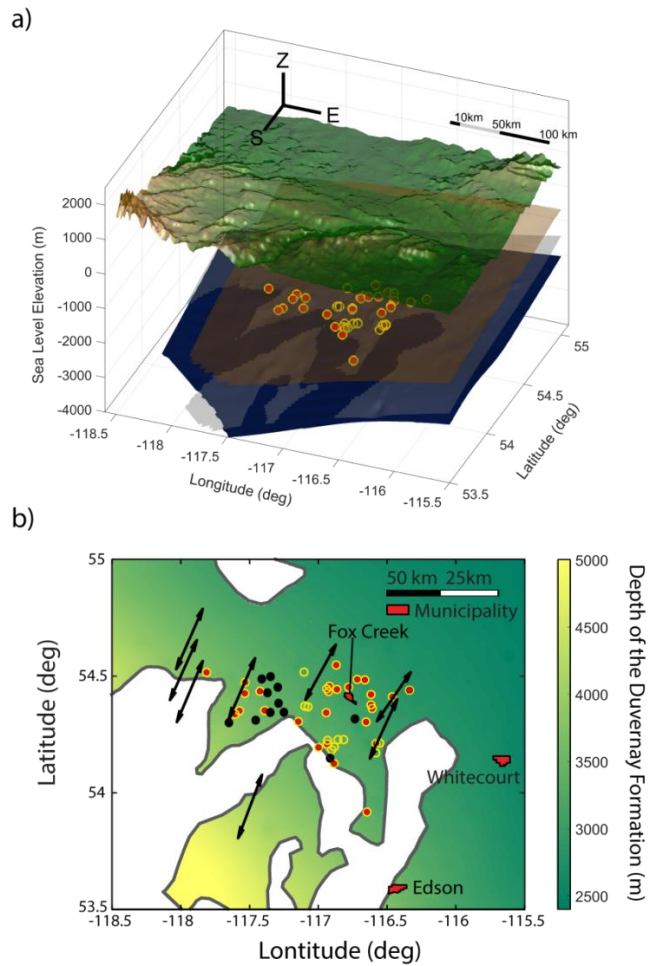

*Fig. 2 The measurements utilized for this study. The surfaces rendered from top to bottom include 30-m resolution shuttle radar topography, and the Mississippian Unconformity (Branscombe et al., 2018), and the tops of the Duvernay Formation extracted from a commercial database and the Neoproterozoic Basement as interpolated from (Peterson, 2017). Red dots and yellow circles indicate the positions of $S_h$ and $P_P$ measurements respectively and many of which are obtained from the same transient pressure record. b) The top-down overview of the study region and the measurement points. Red dots and yellow circle show the locations of $S_h$ and Pp measurements as in a). The black arrows represent the $S_H$ azimuth reveavled by borehole failures intepreted from image logs located within the study area, reported in Shen et al., (2018b). The solid black circles indicate the epicentral positions of suspected induced events (Schultz et al., 2016) as shown in Fig. 1. White areas indicate disruption of the Duvernay Formation by the coeval Leduc Reefs.*

These data are already publicly available (Shen et al., 2018a, b), together with detailed descriptions of the methodologies used to extract this information. Hereafter for brevity we shorten these reports' citations to SA and SB, respectively. We further attempt here to constrain $S_H$ magnitudes utilizing the measurements recorded in SB and focal mechanism solutions for the earthquakes reported within our study area. To avoid confusion, here 'depth' $z$ refers to the true vertical depth (TVD) defined as the vertical distance from the measurement point in the Earth to the drilling rig's Kelly Bushing reference as corrected by any borehole deviation. Elevations $h$ are measured in meters relative to sea level with positive and negative values representing heights respectively above and below sea level. We use the compression positive sign convention that compressive stress, pore pressure, and rock strength all have positive values. Fig. 2a and b shows the positions of the various measurements relative to the epicentres of the major induced earthquakes and important stratigraphic markers.

## 2.1  $S_H$ Orientations $\phi$

We employ the orientations of BOs (Bell and Gough, 1979) and DITFs (Aadnoy, 1990; Aadnoy and Bell, 1998; Brudy and Zoback, 1999) reported in SB the spring-lines of which strike, respectively, perpendicular and parallel to the $S_H$ azimuth $\phi$ (Schmitt et al., 2012; Zoback, 2007; Zoback et al., 2003). The model developed here uses a two-step process in which the statistics of the directions from a given borehole are first defined with these results then spatially interpolated to estimate the stress direction across the study area.

We employed both oriented ultrasonic borehole televiewer and micro-resistivity imaging logs in the analysis; with examples of typical ultrasonic image log data given in Fig. 3. Values of $\phi$ were obtained in the analysis of over 2000 m of borehole image logs taken from 20 boreholes both within (8 boreholes, see Fig. 2b) and close to the study area. The depth extents and directions are tabulated in SB according to quality control assessment criteria described in SA, the analysis here includes four additional logs over those that were available for SB, these new data and a summary of the SB results are included in the supplementary material with WSM quality ranks calculated for reference. It is worth noting that sections of the boreholes interpreted in SB deviate significantly from vertical; in the analysis here, however, almost all log information used is from nearly vertical boreholes with any data obtained at inclinations more than 30° excluded. Clear images of BOs or DITFs are scarce in the segments of the Duvernay

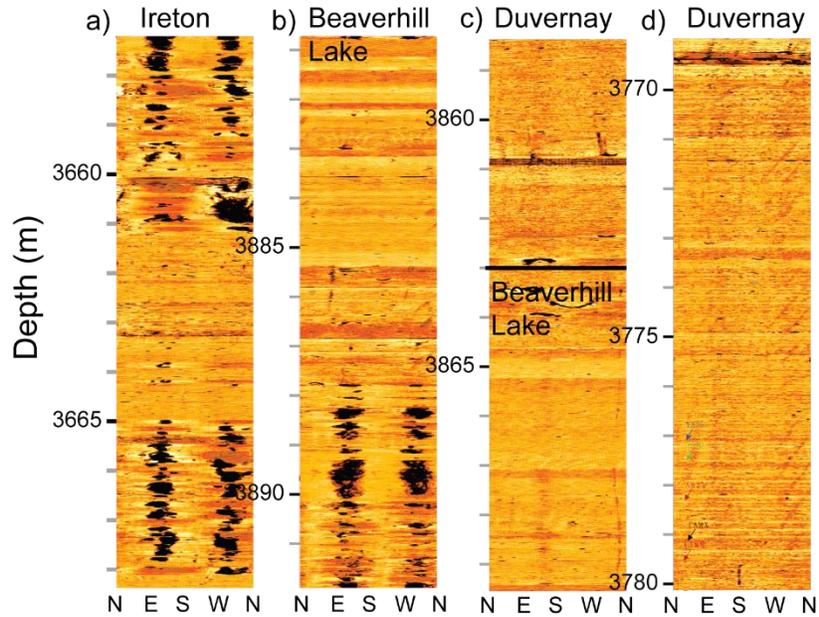

*Fig. 3 Examples of depth segments of high-quality ultrasonic amplitude reflectivity image log data from well UWI 12-11-64-27W5 showing zones with (a and, (b BOs, and without (c and, (d BOs.*

Formation and most recorded stress orientations in SB are from the overlying formations. Interpolation of these directions using an inverse distance weight scheme described below provides the final stress direction map.

The final value of $\phi$ reported at the location of each borehole is the weighted average of its the set of $\phi$ observations as colour-coded in Fig. 4a. This is accomplished using available CircStat software toolbox (Berens, 2009) that calculates the average of a set of complex numbers in Euler form with phase angles and amplitudes equal to the observed $\phi$ and their quality index weights, respectively. This averaging procedure was further adapted to account for the fact that $\phi$ exists only over the azimuth range of $0° \leq \phi \leq 180°$ sufficiently describes the $S_H$ directions by simply adding into the data suite each value's appropriately weighted symmetric counterpart at azimuth $\phi' = \phi + \pi$. The final direction for that data set occurs at the orientation with the maximum probability of the corresponding von Mise's circular probability distribution (Mardia and Jupp, 1999). We illustrate by application to the entire final set of average $\phi$, including those wellbores located outside of our study region, as shown in an equal area rose diagram of the dispersion (Fig. 4b) and the corresponding von Mise's probability distribution (Fig. 4c) with the peak at $\phi = 43°$.

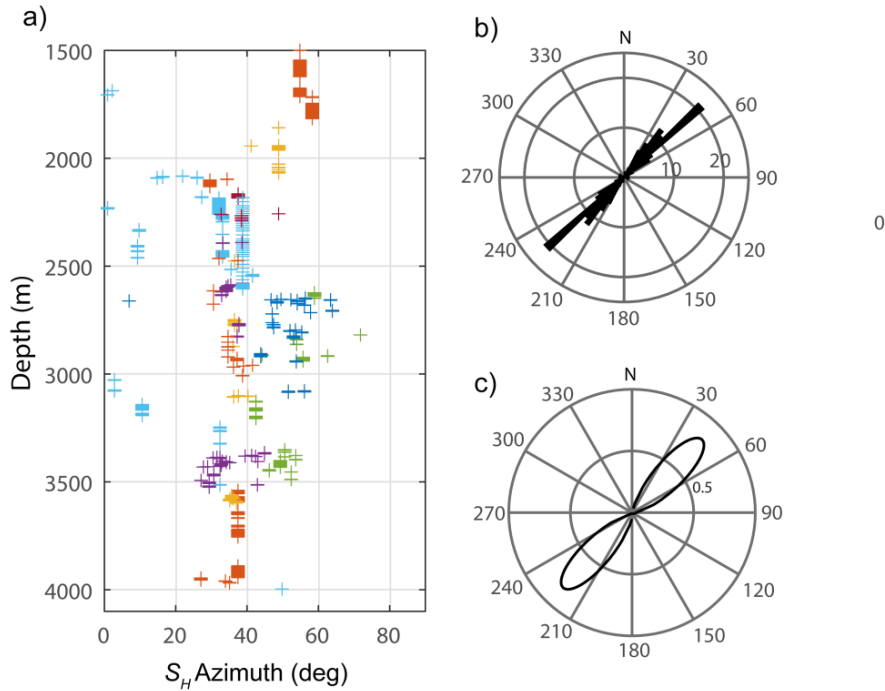

*Fig. 4 a) The complete set of 20 observations of $\phi$, reported in SB, plotted versus depth and colour-coded by each borehole. b) Rose diagram histogram distribution of $S_h$ orientation measurements $\phi$ for the entire set of final average $\phi$ for all the boreholes (including those reported in WSM) used. c) The von Mises circular distribution for the full set of azimuths from a) with the maximum occurring at $\phi = 43°$.*

Many researchers have attempted to develop maps that highlight areal trends in $\phi$ by calculating spatial averages from scattered individual $\phi$ determinations using a variety of averaging methods that have been applied at regional or global scales using earthquake focal mechanisms (Hardebeck and Hauksson, 2001) or stress indicators (Heidbach et al., 2010; Reiter et al., 2014). Given the sparsity of data, these methods usually search over areas with dimensions of 100's of km in order to determine a set of data for averaging. The situation here differs somewhat here that the total study area is more confined with areal dimensions significantly less than the wavelengths of these larger scale studies. For this reason, we take a different approach in which the borehole averaged $\phi$ values are interpolated using an adaptation of the Inverse Distance Weighting (IDW) method with the results shown in Fig 5a. In order to estimate the $\phi$ at a given point in the map, each neighbouring observation '$i$' is weighted $W_i$ according to the ratio of its assigned quality index $Q_i$ to its distance $D_i$

$$W_i = Q_i/D_i \qquad (2)$$

We note that although one must assign a low WSM quality ranking to these BOs and DITF as they are discontinuous, within a given borehole the orientations generally do remain consistent lending more confidence than the formal quality ranking might suggest. That said, the quality of the image logs also varied, and to account for this a numerical quality weight $Q_i$ was assigned to each average $\phi$ semi-quantitatively (see supplementary material). $Q_i \equiv 1$ only for those clear image logs displaying long BO and DITF. $Q_i \equiv 0.75$ for lower quality image logs that still exhibited clear BOs or DITFs that were either of limited extent or perturbed by natural fractures resulting in greater estimated $\phi$ uncertainties. Finally, $Q_i \equiv 0.25$ for those values provided by the WSM primarily because these measures were beyond our control; we do not intend that this necessarily reflects those earlier measurements validity.

Essentially, to obtain a representative value of $\phi$ at a given point in the map the algorithm searches over circles of increasingly larger radii until the sum of all of the $Q_i$ within the circle reaches a threshold value of 3. Once this threshold is exceeded, the IDW weighted average $\phi$ is then calculated using CircStat. The search radius of each of the prediction points in our study area are shown in Fig. 5b. In order to avoid a 'bulls-eye' effect in the averaging from proximate data points, a minimum distance $D_i = 2$ km is enforced although this situation is rare as the distances between the boreholes are mostly larger. The ratio between the weighted standard deviation of measurements $\sigma$ within the search radius to the sum of the corresponding weights

$$A = \frac{\sigma}{\sum W_i} \qquad (3)$$

This confidence measure $A$ (Fig. 5c) accounts for the consistency, the quality, and the proximity of the input measurements relative to the prediction point. Points with high confidence have $A \to 0$. The confidence of IDW $\phi$ is poor in the SW due to conflicting stress directions at some of the boreholes.

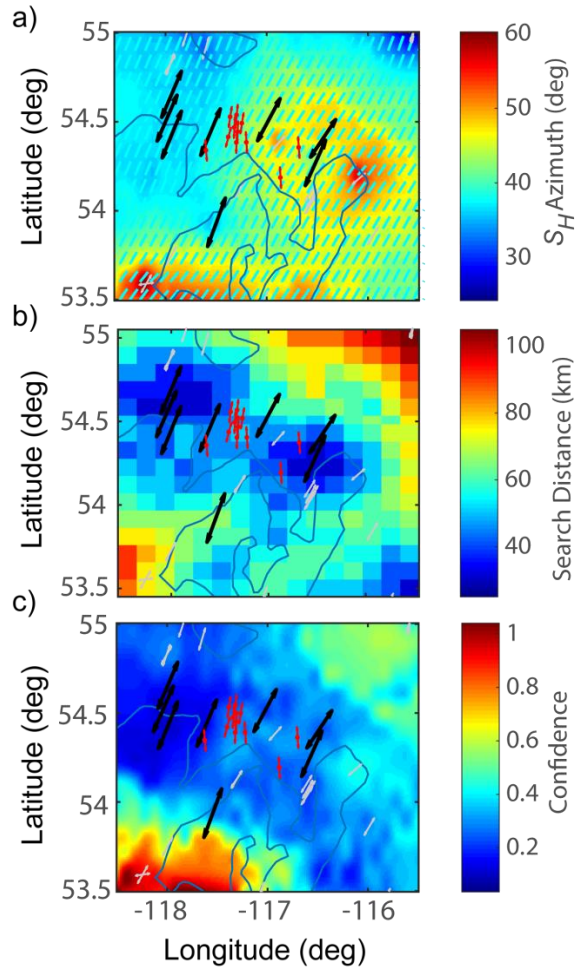

*Fig. 5 a) Map of averaged $S_H$ azimuths $\phi$ determined using IDW. Black arrows represent stress orientation measurements reported by SB. Small white arrows represent those from WSM. The cyan dashes represent directly the modelled $\phi$ at grid points in the study area while the background colour is the smoothed model for $\phi$ over the area. The red arrows and dots show the strikes of the faulting planes revealed by earthquakes' focal mechanism solutions and the epicentral locations. b) The search radius of each prediction point for IDW. c) Confidence in the $S_H$ azimuth prediction. Values close to zero represent high confidence and values close to one stand for low confidence.*

## 2.2 Vertical Stress $S_V$

It is often assumed and generally accepted that the vertical stress $S_V$ is one of the principal direction of the subsurface stress tensor at sufficient depth and that its magnitude is given by overburden pressure due to the weight of the overlying rock mass with density $\rho(z)$. We incorporate 1125 digital $\rho(z)$ logs, with 694 located directly within the study area (Fig. 6e), through the IHS$^{TM}$ digital log database. However, some issues require careful consideration in its practical application. First, the data quality in these logs is often suspect when for example the sensor loses contact with the formation rock in rugose boreholes. Specific procedures used to overcome noise and to incorporate uncertainty from such data were implemented and described in detail in SA. Second, the $\rho(z)$ logs are often only collected over intervals immediately surrounding the zone of economic interest. Measured densities in the topmost 500 m are rare. In order to overcome these restrictions, a 3D density model was constructed by kriging over constant elevations. These calculations were implemented with the adaptation of mGstat (Hansen, 2004), a Matlab$^{TM}$ wrapper for the geostatistical modelling program Gstat (Pebesma and Wesseling, 1998), and variogramfit by Schwanghard (2010) with details provided in SA.

Another issue in the development of a model for SV is that the surface topography varies across the study area. It is well known that topography perturbs both the stress directions and magnitudes (McTigue and Mei, 1981; Miller and Dunne, 1996; Savage and Swolfs, 1986; Schmitt and Li, 1993 ), particularly near the surface. We adapted the 3D Green's function methods (GFM) developed by Liu and Zoback (1992) who needed to account for severe topographic expression near the Cajon Pass scientific drilling project.

Briefly, we separate the crust volume within the study area into a lower elastic halfspace overlain by a layer whose topmost surface matches the topography. The method uses as Green's functions the solutions to the Boussinesq and Cerruti problems that, respectively, give the stress states induced by applying a vertical or a horizontal point load onto the surface of an elastic halfspace. The topographic load stresses are then given as a convolution of these Green's functions with the topographic loads expected at the top of the halfspace. The vertical and horizontal loads applied to the halfspace are determined from the densities and thicknesses of the upper section.

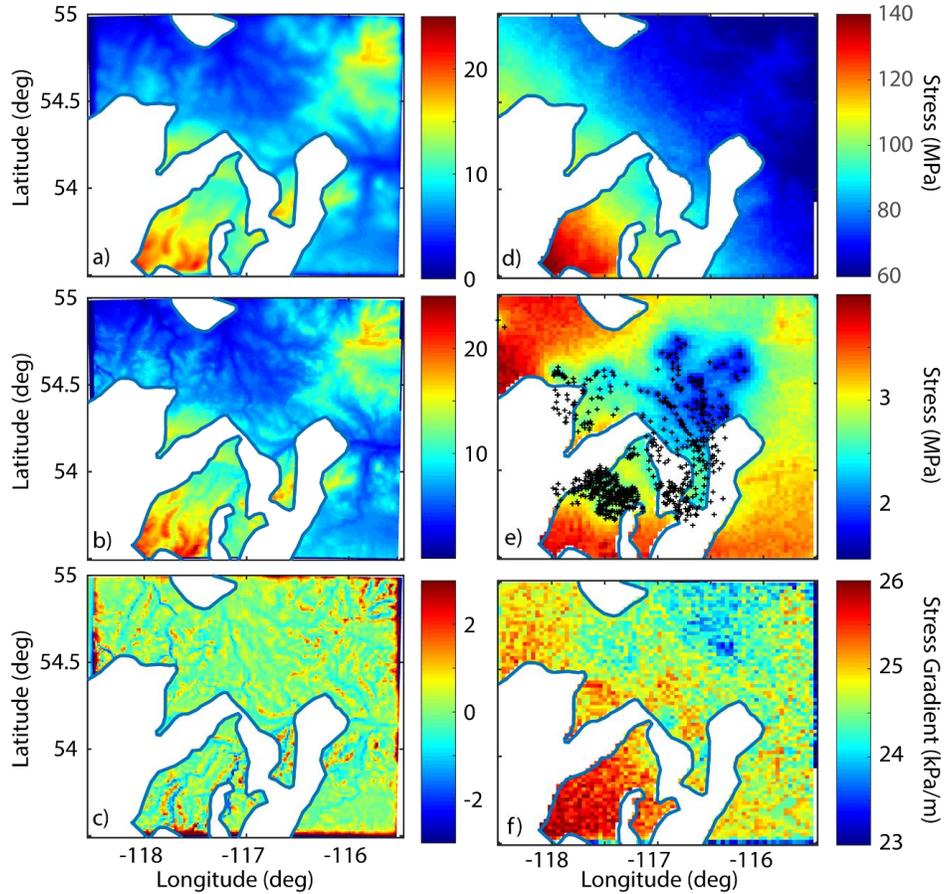

*Fig. 6 a) The lithostatic pressures of the upper section of the model. b) The topographic impact on the $S_V$ at the tops of the Duvernay Formation evaluated through the GFM. c) The difference between the topographic impact and lithostatic pressures of the upper section of the model. d) The vertical stress $S_V$ mapped into the middle depth point of the Duvernay Formation from the elevation corrected model and e) its corresponding uncertainty. The black crosses represent the wellbores with density logs. f) The stress to depth ratio for $S_V$.*

An elevation $h = 558$ m that is ~20 meters below the lowest surface elevation over the study area separates the upper section and lower halfspace. The summation of the topographic impact from the upper section and the lithostatic pressures of the lower halfspace is an approximation of the actual $S_V$ (Liu and Zoback, 1992). To overcome the limits of the sparse density logging information at shallow depth, we gathered the boreholes with density logs with elevation 100 meters below the top of lower halfspace (elevation > 458m). The average densities of these boreholes are then adapted to make the 2D near surface topography density map through kriging. The lithostatic pressures of the lower halfspace are calculated using the 3D density volume.

Fig. 6a shows the topographic impact component of the $S_V$ at the tops of the Duvernay Formation with a comparison to the lithostatic pressures (Fig. 6b) and a noticeable difference is observed (Fig. 6c) The GFM suppresses the short-wavelength topographical features while the influence of the long-wavelength, regional topographical variations remain at the depth of the Duvernay Formation. Evaluating the $S_V$ only considering the densities of rocks can result in an error of up to 1.5 MPa at the depth of the Duvernay Formation, at vicinities where topography varies rapidly. Gaps in the boundaries of the figure are the artifacts caused by the conversion of the input graticular coordinates to a Cartesian coordinate for faster calculation of distances.

We project the final modelled $S_V$ magnitude (Fig. 6d) and its uncertainty as determined through the propagation of errors (Fig. 6e) onto the tops of the Duvernay Formation. The uncertainties for $S_V$ range from 2 MPa to 5 MPa and primarily reflect the local concentration of available density log data (see Fig. 6e). The secant gradient of $S_V$ (Fig. 6f), which is merely the ratio of the calculated value of $S_V$ to the depth used widely in industry to allow for simple estimation of stress, is roughly 24-26 kPa/m.

### 2.3 *Minimum Horizontal Compression $S_h$*

We interpreted 30 selected well pressure tests, variously referred to as micro-frac, mini-frac, or in this case, Diagnostic Fracture Injection Tests (DFITs) in order to provide estimates of $S_h$ magnitudes with the details of the methods employed provided in SA and the results in SB. All of the measurements analyzed for this paper were carried out within the Duvernay Formation, but we also include in the supplementary material some earlier values from the overlying Cretaceous section provided by Woodland and Bell (1989). Briefly, the raw data used is from time series of the wellbore pressure $P_w(t)$ measured during testing. A series of repeated pressurizations and fracture re-openings with controlled pressure declines are preferred for stress measurement (Bell et al., 1994; Schmitt and Haimson, 2016), but standard industry practice employs only one single pressurization and decline cycle in tests that can run for many days in the low permeability formations. Typically, an interval of the borehole is sealed using packers. This interval is then rapidly pressurized until $P_w(t)$ drops indicating the creation of a fracture in the formation. Pumping to the formation may continue in order to extend the fracture into the formation further. This fracture is presumed to grow in the direction of $S_H$ and perpendicular to the minimum horizontal compression. When deemed appropriate by the operator pumping ceases and the interval is 'shut-in' allowing $P_w(t)$ to evolve as fluid diffuses from the induced fracture that is assumed to still be open into the formation. $P_w(t)$ eventually decays to the $S_h$ magnitude at which point the fracture closes and the $P_w(t)$ decline behaviour changes. Hence, finding this fracture closure pressure $P_{fc}$ at this point of departure allows $S_h$ to be estimated.

Four different approaches, described in detail in SA, were used on each available $P_w(t)$ record providing a range of values allowing a level of uncertainty to be assessed. Due to the lack of information we have not considered instrumental error as being important as typical errors for pressure gauges used in such studies are small relative to the pressures observed, and further, this information is not generally provided in the records available to us. More serious deviations could come from the fact that pressures within the 'closed' interval may, in many instances, be measured at the surface and as such the values reported at depth here may be corrected for hydrostatic head differences that are sensitive to the pressure and temperature variations of the borehole fluid. We further assume that the TVD values provided are correct. In a small number of cases only already interpreted values of $P_{fc}$ were available as $P_w(t)$ was not provided by the operator; these are arbitrarily assigned a higher uncertainty of 5% as we were unable to assess the original quality of the data.

When plotted with respect to depth, irrespective of spatial location, the $S_h$ magnitudes generally increase (Fig. 7a) although with some scatter may reflect the heterogeneity of the stress field. Linear regression of the $S_h$ - $z$ depth relation gives directly

$$S_h(z) = 32.1 \pm 3.1 \frac{kPa}{m} z - 41.8 \pm 10.2 \; MPa \tag{4}$$

to 95% confidence.

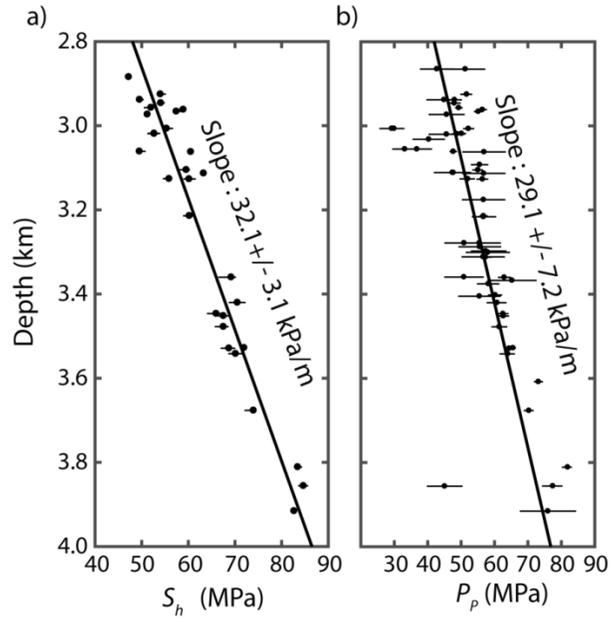

*Fig. 7 Magnitudes of a) $S_h$ and b) $P_p$ versus depth with error bars representing uncertainties. Black lines show the best linear fit for each case.*

The regression analysis above provides some capacity for $S_h$ prediction, but it cannot include any spatial variations in the stress field. We account for such variability in a second model for $S_h$ established by kriging of the available measurements. In order to do this, before interpolation, we first removed the depth-dependent trend by shifting the observed $S_h$ to the same depth using the tangent gradient $\Delta S_h/\Delta z = 32.1$ kPa/m $\pm$ 3.1 kPa/m from Eqn. 4. These interpolated data were then corrected down to the depth at the top of the Duvernay Formation to provide the spatially varying model $S_h(x,y)$ shown in Fig. 8a. The spatial model for $S_h$ with its associated uncertainty (Fig. 8b) is available in the supplementary materials.

The differences between this interpolated model and the more straightforward linear trend of Eqn. 4 (Fig. 8c) show a maximum spatial difference of ~10 MPa from the highest to lowest values. These differences do not appear random. For example, the modelled values exceed those of the linear trend over an E-W band from about -117° to -116.5° near the center of the study area. We compare the $S_h$ and $S_V$ by subtracting model $S_v$ (Fig. 8d) from the corresponding model $S_h$ (see Fig. 8a). The result (see Fig. 8d) suggests that the difference between the two decreases with depth.

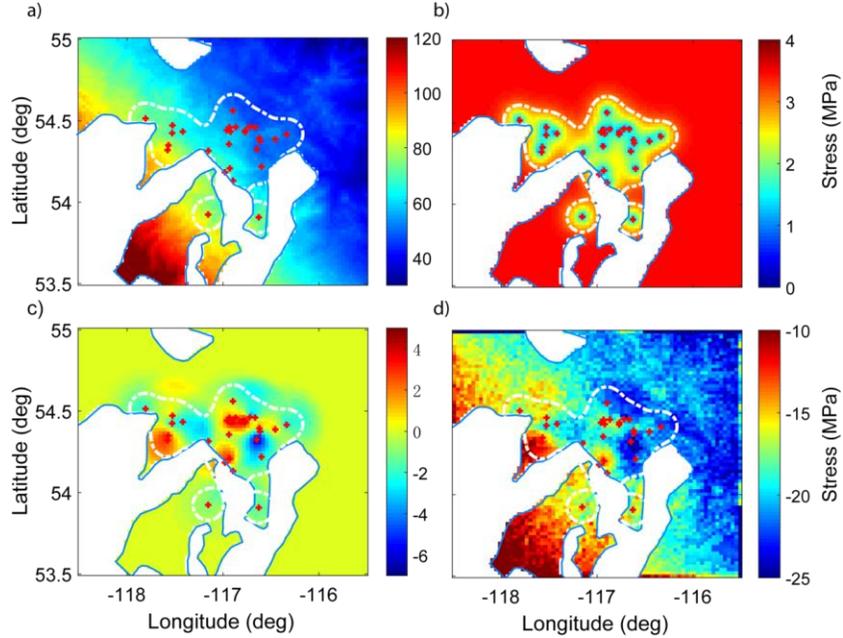

*Fig. 8. Projections to the top of the Duvernay Formation of a) the modelled $S_h$ value. b) The uncertainty of the modelled $S_h$. c) the spatial perturbation of $S_h$ calculated by subtracting the modelled $S_h$ value with the linear trend of Eqn. 4. d) The difference between $S_h$ and $S_v$. In all panels the white dashed line represents the contour with an uncertainty of 3 MPa. Empty spaces indicate the engulfed Leduc reefs. Red crosses show the locations of the $S_h$ observations.*

## 2.4  Formation Pore Pressure $P_P$

We also include 57 measurements of fluid pressure as indicated by various techniques including i) asymptotic extrapolations of the decline in $P_w(t)$ after fracture closure in the tests just described, ii) static pressure gauge tests, and iii) flow/build up tests (see SA for additional details). We note that these are values that are obtained usually within a section of the cased but perforated borehole after relatively long periods (sometimes days) once the cased well has been opened. They represent the equilibrium pressure or its estimate measured within the borehole that is presumed to indicate a uniform virgin pore fluid pressure within the rock mass.

Regression of these observed $P_P$ with depth (Fig. 7b) gives a trend

$$P_P(z) = 29.1 \pm 7.2 \frac{kPa}{m} z - 39.6 \pm 23.4 \; MPa \tag{5}$$

with a tangent slope $\Delta P_P/\Delta z$ = 29.1 kPa/m ± 7.2 kPa/m. Using the same methodology as applied in the development of the $S_h(x,y)$ map above, a spatially interpolated map of $P_P$ was calculated by kriging (Fig.

9a) with these interpolated values provided in the supplementary material. Despite having a larger number of measurement points, the uncertainty in $P_P$ (Fig. 9b) is nearly double that of $S_h$. Such uncertainty primarily arises from the errors inherent in attempting to extend the slowly dissipating $P_w(t)$ to infinite time, and the greater uncertainties assessed the static values.

The observed and modelled $P_P$ significantly exceeds what is expected for a normal hydrostatic gradient, assuming a groundwater specific weight of 9.8 kN/m$^3$ (Fig. 9d). However, there appear to be coherent zones where the modelled $P_P$ is significantly diminished relative to the trend of Eqn. 5 (Fig. 9c) that may be related to the contacts at the Duvernay Formation and the Leduc reefs.

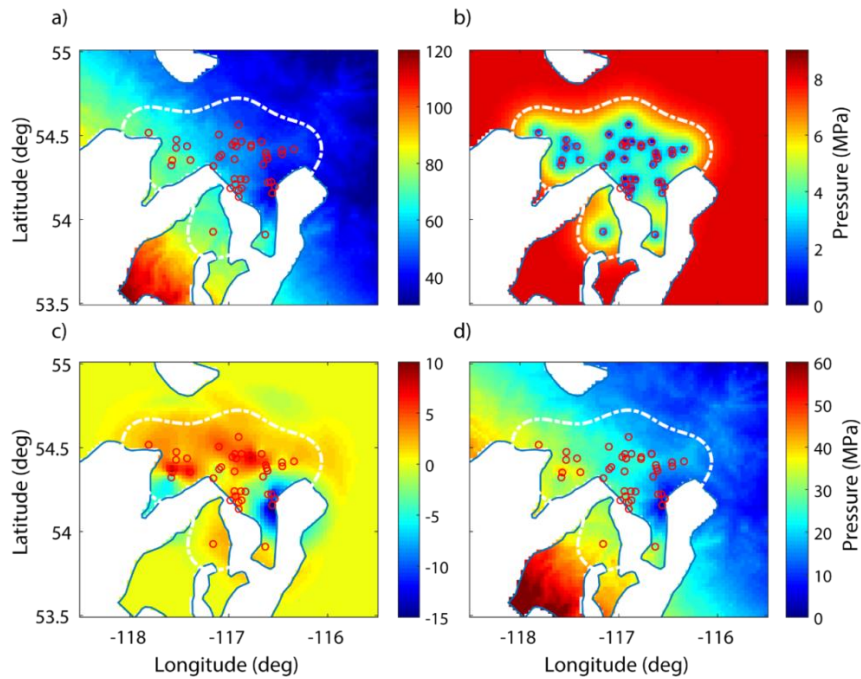

*Fig. 9. 3D projections to the top of the Duvernay Formation of a) the modelled $P_P$ value. b) The uncertainty of the modelled $P_P$. c) The spatial perturbation of $P_P$ calculated by subtracting the modelled $P_P$ value with the linear trend of Eqn. 5. d) The difference between $P_P$ and hydrostatic fluid pressures. In all panels, the white dashed line represents the contour with an uncertainty of 7 MPa. Empty spaces indicate the engulfed Leduc Reefs. Red circles show the locations of the $P_P$ observations.*

## 2.5   Greatest Horizontal Compression $S_H$

Unlike $S_h$, the determination of the magnitudes of $S_H$ remains challenging as the direct measurement from deep boreholes is technically difficult and cost prohibitive. In SA, assessment of $S_H$ is provided using the formulation had been previously applied for the KTB borehole (Barton et al., 1988), assuming the negligible difference between the formation $P_P$ and drilling fluid pressures. However, such calculated $S_H$ values are subject to significant uncertainties as results of the mostly unknown input $P_p$, $S_h$ and rock strength at the depths of the observed BO, which are often more than one-thousand meters shallower than the depths of the Duvernay Formation. We did initially attempt to directly constrain the $S_H$ magnitudes exploiting observed DITF and BO widths from the segments of image logs from the Duvernay Formation. However, unacceptably high uncertainties primarily due to the high values of $P_p$ led us to abandon this approach.

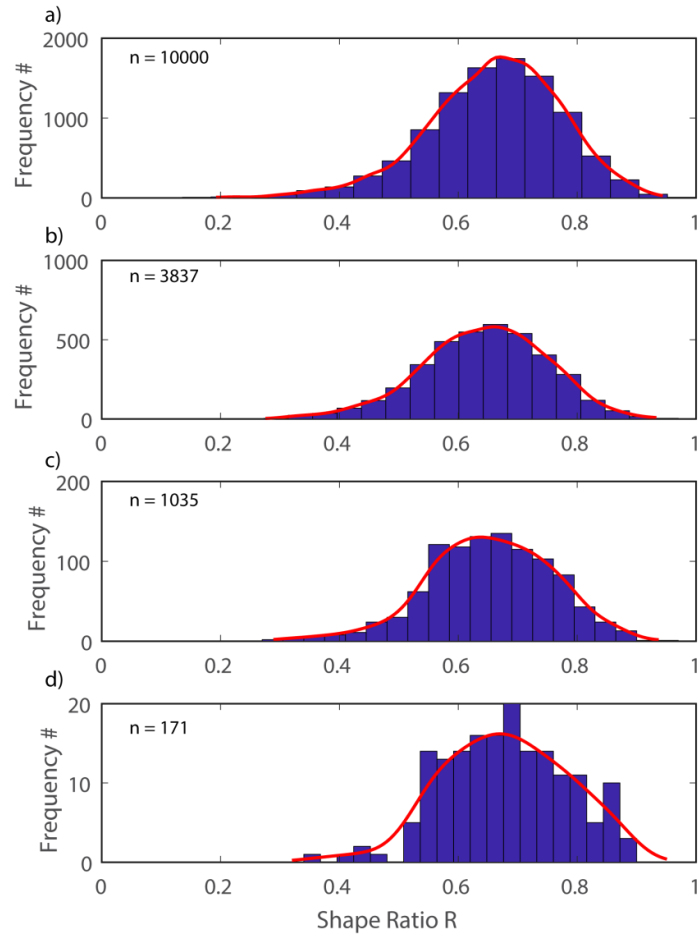

*Fig. 10 histogram of the shape-ratio R from stress inversion with red line represents the nonparametric kernel-smoothing distribution fit (Bowman and Azzalini, 1997) that include a) all inversion results regardless of the $\sigma_2$ dip; or only those results with $\sigma_2$ dip angles b) greater than 80° c) greater than 85°; d) greater than 88°*

Alternatively, the $S_H$ magnitudes were constrained by the inversion of the induced earthquakes focal mechanisms in a manner similar to that employed recently by Terakawa and Hauksson (2018). The methodologies for stress inversion are well developed (Gephart, 1990; Jones, 1988; Kastrup et al., 2004; Montone et al., 2004), with a brief synopsis included in the appendix. This method provides a single measure of the principal stress magnitudes $\sigma_1 \geq \sigma_2 \geq \sigma_3$ via a shape factor $R$:

$$R = \frac{\sigma_1 - \sigma_2}{\sigma_1 - \sigma_3} \tag{6}$$

Notably, one disadvantage of such approaches is that they presume the uniformity of the regional stress field throughout the volume of crust studied regardless of position, depth, or lithology. As such, the ability to obtain an independent measure of a local stress magnitude directly, as might be hoped for from a borehole, is sacrificed. However, stress field homogeneity may be a reasonable assumption given the consistency in $\phi$ and the behaviour of $S_h$ with depth (see Eqn. 4).

The ambiguity between the fault and the auxiliary plane is problematic when focal mechanisms are used. Following the Mohr-Coloumb frictional faulting criteria (Eqn. 1), Lund and Slunga (1999) sought the correct plane by testing which was the least stable and hence expected to slip. Vavryčuk (2014) iteratively applied this criterion to Michael's (1984) inversion (see Eqn. 7); this necessitates a range of values of $\mu$ to be input in addition to the slip and fault plane estimates provided from the suite of observed focal mechanisms. In this study, we used the Varycuk's (2014) stress inversion program in which the focal mechanism strikes, rakes, and dips (see supplementary material) for the ensemble of earthquakes (see Fig. 1b) are used, to both provide an $R$ value and to better determine which nodal planes best represent the faults. A broad range of $0.2 \leq \mu \leq 1.2$ for friction is used given there are no laboratory constraints on rock friction. Mean random errors of 10° were assigned to the input values for the $n_k$ taken from the focal mechanisms.

Ten-thousand realizations with these differing parameters were inverted from the 11 earthquake focal mechanism solutions (see Fig. 1b and supplementary material) producing a distribution of $R$ values (Fig.

10a). The Probability Density Function (PDF) fit to this distribution using nonparametric kernel-smoothing distribution fit indicates the most probable value for *R* is 0.67 with 90% probability of falling in the range 0.46 < *R* < 0.84.

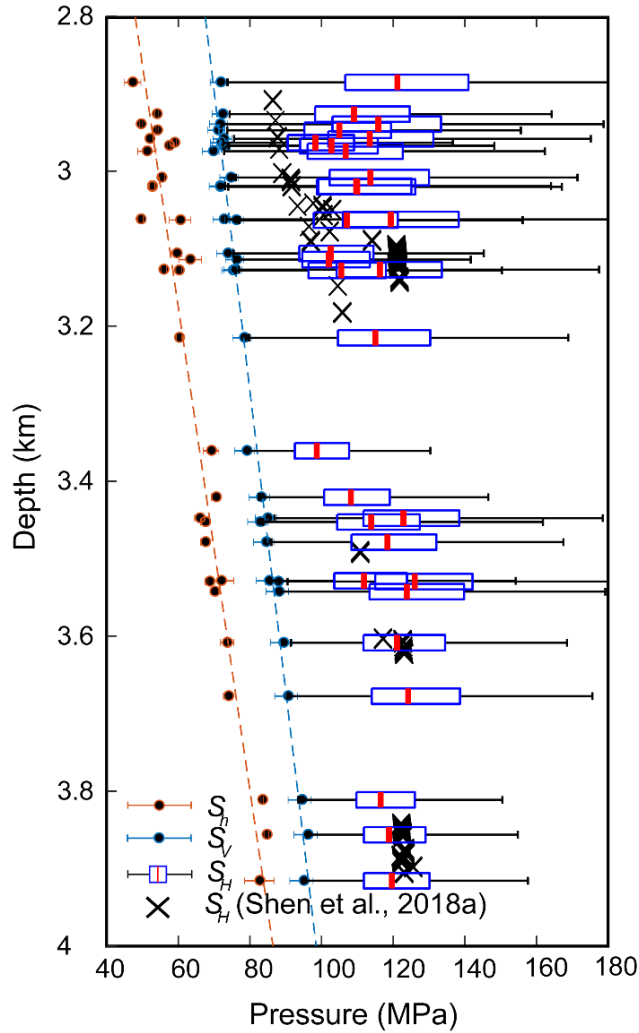

*Fig. 11 $S_h$ measurements and calculated $S_H$ from SB with $S_V$ and $S_H$ modelled in this study at the location of the $S_h$ measurements. The Whisker-Box plot of $S_H$ denotes the range of possible values of $S_H$. The blue boxes confine 25th and 75th percentile of the CDF of $S_H$, and red dashes present the most probable value of $S_H$. Whiskers represent the minimum and maximum values of $S_H$. Black crosses represent the magnitudes of $S_H$ initially estimated by SA.*

The statistical distribution of *R* is further utilized to calculate the probability distribution of $S_H$ through a Monte-Carlo simulation using 5000 realizations assuming that $S_h$ and $S_V$ are coaxial to $\sigma_3$ *and* $\sigma_2$ respectively, in accordance with the strike-slip focal mechanisms. More discussion on the justification of such assumption is provided in section 3.3. We account for the uncertainties of $S_h$ ($\sigma_3$) and $S_V$ ($\sigma_2$) by assuming a uniform distribution of random noise based on their assessed observational uncertainties (see section 2.2 and 2.3). Fig. 11 shows the constrained $S_H$ values and probability distributions at the locations of each $S_h$ observation with the statistical distribution of $R$ shown in Fig. 10a.

## 3      Discussion

### 3.1    $S_H$ Directions

The map for the $S_H$ direction (see Fig. 5a) shows that $\phi$ is reasonably uniform across the study area and consistent with the earlier studies indicating the general NE-SW compression that appears to persist within the Western Canada Sedimentary Basin (Bell and Gough, 1979; Reiter et al., 2014). The N43°E average $\phi$ determined here is close to the ~N45°E found using a broader search area with a radius of ~500 – 1000 km (Reiter et al., 2014). Stress inversion performed on the focal mechanism solutions of the earthquakes within the region (see Fig. 1) yields a most probable $\phi$ of ~N55° E (Fig. 12) with a probability distribution varying from N40° E – N60° E, slightly higher than our predicted $\phi$, using solely borehole observations, at the locations of the earthquakes between N39°E to N45°E (see Fig. 5a and Fig. 12). Irregular geological features, in particular, the Leduc reefs hypothesized to have nucleated along faults, do not appear to perturb the stress directions. Moreover, we do not observe stress rotation at vicinities near the epicentres of the reported earthquakes.

The predominantly strike-slip focal mechanisms observed for the Fox Creek earthquakes (Bao and Eaton, 2016; Schultz et al., 2017; Wang et al., 2016; Zhang et al., 2016) are consistent with fault slip on vertical planes that strike early N-S. $S_H$, with $\phi$ ~39-45° at the regions of near the epicentres of these earthquakes, acts to the inferred fault plane that in turn leads to a frictional angle $\psi = 14 - 52°$. These friction angles correspond to reasonable estimates for the coefficient of friction $\mu$ ~ 0.3 or higher. If the regional average (~43°) or the longer-wavelength modelling result (~45°) are to be used (Reiter et al., 2014), the frictional angles are subsequently decreased to $\psi$ ~ 0 – 7° with correspondingly very low coefficient of frictions < 0.1. Though, care must be taken with this simplified analysis as it does not fully account for the possible combinations of fluid pressures and stresses resolved onto the fault plane. A more complete analysis is currently underway.

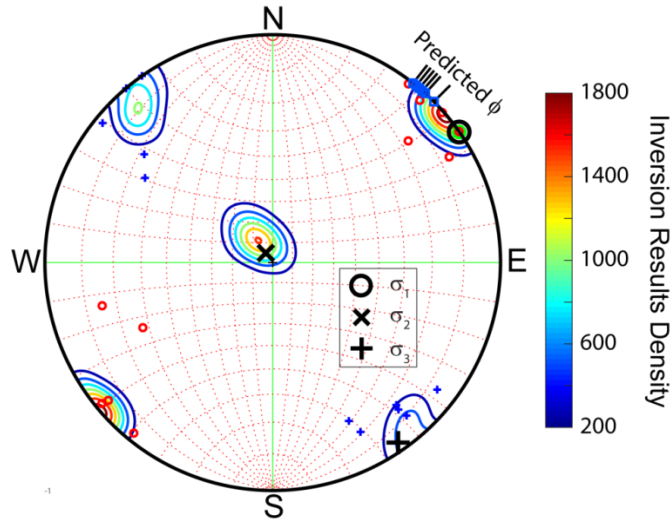

*Fig. 12 Axes of $\sigma_1, \sigma_2, \sigma_3$ from the recorded earthquake's focal mechanism solutions and stress inversion. Contours represent densities of $\sigma_1, \sigma_2, \sigma_3$ solutions from stress inversion. Red circles and small blue crosses represent the P/T axes from the input focal mechanism solutions. Black symbols show the most probable orientation of the in-situ stress by stress inversion. Blue squares and lines indicate the orientations of $S_H$ predicted from the stress model at each of the earthquake's epicentral locations shown in Fig. 5*

### 3.2  $P_P$ and $S_h$ Gradients

The observed $P_P$ all substantially exceed those anticipated from a normal hydrostatic (water) gradient of ~9.8 kPa/m that assumes continuous hydraulic conductivity from depth to the surface; and, indeed, even exceed the expected wellbore fluid pressure $P_w$ with gradients bounded below ~14 kPa/m due to high mud densities reported. Using our data, the 'secant' $P_P$ gradient that is usually given in standard industry practice as the simple ratio of pore pressure to depth ranges from 12.3 kPa/m to 20.3 kPa/m. This mostly agree with the simpler secant gradients to the Duvernay Formation in the areas recently published (Eaton and Schultz, 2018; Lyster et al., 2017) and to values ranging 16 kPa/m to 22 kPa/m more globally from other researchers (Bell and Grasby, 2012; Haug and Bell, 2016; White and Foxall, 2016; Yerkes et al., 1990).

However, these simple secant $P_P$ gradients underestimate the much higher local 'tangent' gradient $\Delta P_P/\Delta z$ ranging between 28.8 kPa/m ± 3.9 kPa/m determined from the linear regression with the depth of the $P_P$ observed within the Duvernay Formation within the study area as indicated by Eqn. 5. This is particularly worth noting as it exceeds the tangent gradient of ~26 kPa/m expected for lithostatic gravitational load $S_V$ within the Paleozoic section. The corresponding $S_h$ secant gradients range from 16.1 kPa to 21.9 kPa that, similarly, is significantly less than the tangent $S_h$ gradient $\Delta S_h/\Delta z$ = 32.4 kPa/m ± 4.2 kPa/m (Eqn. 4). Before this study, workers (McGarr, 1988) have argued that lithostatic pressure provides an appropriate reference stress state (Zang and Stephansson, 2010); and consequently, the higher rate of increase of the pore pressure over the vertical stress observed here is unexpected (Swarbrick and Osborne, 1998). The approach of $P_P \rightarrow S_h$, too, may be meaningful and may suggest a linkage between pore pressures and tectonic loading. Conversely, within the shallower overlying Cretaceous formations, the $P_P/S_h$ ratio is significantly lower Woodland and Bell (1989) although these values may have been compromised by earlier hydrocarbon production (Bell et al., 1994).

## 3.3  *Validity of Stress Inversion and Andersonian Assumption*

We constrain the magnitudes of $S_H$ through stress inversion on the focal mechanism solutions of the earthquakes recorded within our study area and near the depth of the Duvernay Formation. The stress inversion using multiple earthquakes' focal mechanism solutions fundamentally relies on two assumptions affecting its validity: 1. the slip direction and the shear traction on the faulting plane are parallel; and 2. The stress field in the rock mass encompassing these epicentral locations is uniform (Pollard et al., 1993). Assumption #1, alternatively known as the Wallace-Bott hypothesis (Bott, 1959; Wallace, 1951), is generally accepted for mid-upper crustal earthquakes (Célérier et al., 2012).

The validity of assumption #2 is often questioned (Faulkner et al., 2006; Pollard and Segall, 1987 ; Zoback et al., 1989) and it is critical to determine whether it can reasonably apply here particularly given that we do see variations in the stress field. Some researchers have attempted to overcome stress field heterogeneities by dividing their regions into subzones with similar focal mechanisms over which the stress field is assumed consistent (Hardebeck and Michael, 2006; Hicks et al., 2000; Kastrup et al., 2004; Levandowski et al., 2018). The subzones in these studies are usually much larger than our current study area, and although we observe noticeable spatial variations of the stress field, there is no evidence for stress concentration or rotation near the epicentres of the observed earthquakes. Further, the focal mechanisms used in the stress inversions also appear to be reasonably uniform. Few earthquakes were reported in our study area before the recently induced seismicity (Atkinson et al., 2016; Schultz et al., 2017) and this too might suggest that more localized stress concentrations along and at the ends of slip zones are not complicating the stress field. Finally, the good agreement between the stress orientations determined from the borehole image logs with those arising from the inversion further suggests a uniformity of at least the stress directions.

The stress inversion itself cannot provide quantitative stress magnitudes directly. Values of the two principal components must be known in order to calculate the magnitudes of the remaining one. So far, we developed our stress model under the assumption of an Andersonian stress regime in which $S_H$, $S_V$, $S_h$ are equal to $\sigma_1$, $\sigma_2$, $\sigma_3$ respectively. It is questionable whether such an assumption is valid as the stress orientation revealed by stress inversion shows, small but noticeable, deviation from vertical and horizontal axes (see Fig. 12). To test the sensitivity of our $R$ to the deviation of the stress tensor, we plotted the histograms of $R$ from all randomly simulated results (see Fig. 10a), along with $R$ corresponding to stress tensor with $\sigma_2$ deviated 10° (Fig. 10b), 5° (Fig. 10c) and 2° (Fig. 10d) from vertical. The fitted PDFs for each of the plotted histograms are stochastically identical, suggesting that our assessed $R$ is not impacted.

Also, the most probable stress orientation revealed by the stress inversion, honouring the requirement that stress axes are orthogonal to each other, remains nearly Andersonian with a deviation of $\sigma_2$ from vertical of less than 5°. Additionally, the 'uncertainty' of $R$, assessed through stress inversion, reflects the spatial variation of states of crustal stress on top of the mathematical errors. Very generous uncertainties (10° for each component) are given for the input focal mechanisms to account for errors in seismological recordings resulting conservative assessment on the range of $R$ and possibly contribute to the deviation of stress axes. Thus, we conclude with confidence that our assumption of Andersonian stress regime and the subsequently constrained ranges of $S_H$, assuming that $S_h$ and $S_V$ are parallel to $\sigma_1$ and $\sigma_2$, are valid.

It is also worth noting that, though drawbacks of $S_H$ assessed through borehole failures (see section 2.5) results in rather large uncertainties, the ranges of $S_H$ constrained in this study is consistent with the values from SA calculated with borehole observations. However, without detailed rock mechanics and hydrogeological knowledge for the formations where the BOs are observed in SA, we cannot say if such an agreement is a coincidence.

Because of the uniformity to the stress field imposed by the inversion, the depth and spatial variation $S_H$ assessed through this approach is only a reflection of the varying $S_V$ and $S_h$. We cannot produce a spatial map or vertical profile of $R$ or subsequently $S_H$, given the uncertainties of on the depths and focal mechanism solutions of the earthquakes mixed up with spatial variation.

### 3.4 3D Stress Model

The primary purpose of this study is to develop a model that allows for the states of in-situ stress within and near the Duvernay Formation to be quantitatively estimated. For a given input location (in terms of latitude and longitude) and depth, the Matlab$^{TM}$ program, threeD_stress_duvernay, provides the Andersonian stress magnitudes [$S_H$, $S_V$, $S_h$], the pore pressure $P_p$, and the stress orientation $\phi$ with conservative estimates of uncertainty. The model incorporates the 2D interpolated values $\phi$, $S_V$, $S_h$ and $P_P$ of the data used to make the maps shown respectively in Figs. 5, 6, 8 and 9 as well as the datum depth at the tops of the Duvernay Formation (see Fig. 2). These values are laterally refined by cubic spline interpolation based on the latitude and longitude of the point of interest. We assume that $\phi$ remains unchanged with depth. The datum depth values for $S_h$ and $P_P$ are corrected to the depth of interest using the tangent slopes of Eqns. 4 and 5 accounting for the uncertainties. A similar method is used to estimate $S_V$ with conservative tangent lithostatic gradients (23.5 kPa/m – 27.4 kPa/m) based on rock densities from 2400 kg/m$^3$ to 2800 kg/m$^3$. A distribution for $S_H$ is then determined using a number of random Monte-Carlo simulations, with updated values of $S_h$ and $S_V$, through the probability distribution function for $R$ (see Fig. 10a).

One issue with the current model is that all of the $S_h$ and $P_P$ measurements obtained originated from well testing solely within the Duvernay Formation; and our prediction of states of stress nearby relies upon their linear extrapolation from that depth. Stress relaxation, often caused by faulting (Hergert et al., 2015; Vernik and Zoback, 1992) or time-dependent viscous deformation (Sone and Zoback, 2014), could perturb the stress field with time resulting in lower than expected horizontal stress magnitudes (Plumb et al., 1991). Sone and Zoback (2013) reported a series of laboratory experiments describing the viscous deformation of shale and suggested considerate stress relaxation over the geological time. That being said, opposed to the samples tested in their reported work, the Duvernay shale generally consists of low clay averaging at 26% (Hammermaster et al., 2012). Such mineral composition generally results in less creep potential (Sone and Zoback, 2013). Ultrasonic and static strain anisotropy measurements show high stiffness in the direction parallel to layering that is similar to a pore-free quartzite (Ong et al., 2016). As such we do not expect significant stress relaxation within it.

There is no data, to our knowledge, regarding time-dependent behaviour of the surrounding formations. However, we argue that this local extrapolation still adequately predicts the states of stress within a reasonable depth range of about ±200m from the Duvernay Formation). Fig. 13 shows that the linear depth relationship established with $S_h$ measurements exclusively obtained from the Duvernay Formation remains valid when compared with measurements obtained from overlaying formations across the Alberta basin. The gradients of the $S_h$ from these two data sets become smaller at the depth shallower than 2.5 km probably because these shallower measurements are primarily within the lower density Cenozoic siliciclastics. However, the gradients and linearity trend of the Haug and Bell (2016) data below 2.5 km depth is consistent with the present Duvernay Formation measurements. Combined regression of the current data with that below 2.5 km from Haug and Bell (2016) gives a tangent slope of 31.5 kPa/m ± 2.5 kPa/m, close to the gradient of 32.1 kPa/m ± 3.8 kPa from Eqn. 4, which was calculated using exclusively SB data. For context, the shallowest measurement from SB is measured at a depth of 2.9 km. Accounting the uncertainties associated with the Haug and Bell (2016) measurements, we are confident the linear relationship described by Eqn. 4 is applicable to the depths reasonably close to the Duvernay Formation.

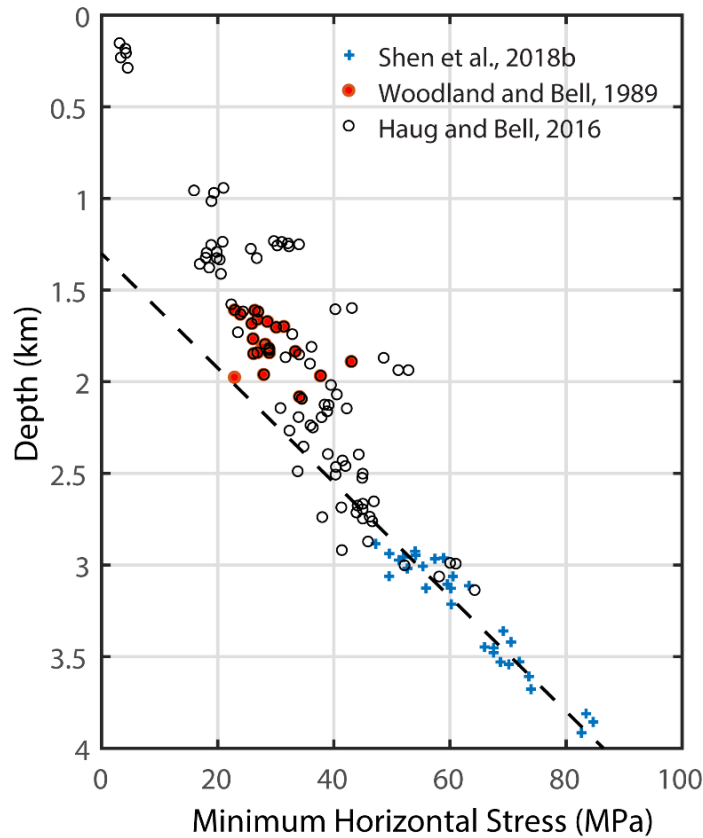

*Fig. 13 $S_h$ measurements reported in Woodland and Bell (1989), Haug and Bell (2016) and SB. The dashed line represents the linear regression results on SB data (see Eqn. 4)*

## 4     Conclusion

We developed a model to predict the full set of stress components describing the Cauchy stress tensors for the states of in-situ stress within the Duvernay Formation, in a locale subject to a number of earthquakes induced by hydraulic fracturing operations. This type of model is only possible because of the density of direct stress measurements from hydrocarbon development boreholes in the study area and to our knowledge, unique.

Transient well testing results provided measurements on the $S_h$ and $P_p$ allowed us to extrapolate for crustal volumes nearby. The large quantity of density logs from oil/gas operations facilitated our efforts of establishing a 3D density model and allowed the calculation of $S_V$. We performed stress inversion on 11 well resolved focal mechanisms from earthquakes recorded in the center of our study region. The stress inversion results supported the assumed Andersonian stress regime and enabled estimation of $S_H$.

The compiled stress measurements further enabled a predictive model allowing assessment of states of stress within our study area, at depths reasonably close to the Duvernay Formation. Full set of the components describing the Cauchy tensor of the in-situ stress can be obtained, through the provided program threeD_stress_ duvernay, for the given location and depth within our study area. Such a model is useful for analyzing the source mechanism of the recent nearby earthquakes. As we recognize that the model is imperfect, we further attempt to include estimates of uncertainty. The heterogeneity of the stress gradient and $S_H$, though not directly accounted in our model, is captured within our uncertainty ranges.

## 5      *Appendix*
*Synopsis for inversion of earthquake's focal mechnism solutions for constraints on $S_H$*

The ensemble of faults with different known orientations used in the inversion are all presumed to be subject to a uniform stress field $\sigma_{ij}$. A traction vector $T_k$ calculated from this stress field acts on each fault plane; and it is resolved into a normal $\sigma_k$ and a shear $\tau_k$ component, with the direction of the latter described by unit vector $N_k = N_{k1}x_1 + N_{k2}x_2 + N_{k3}x_3$. Angelier (1979) obtained the former from the faults' dips and strikes and the latter from the corresponding slickenslide rakes $s_k = s_{k1}x_1 + s_{k2}x_2 + s_{k3}x_3$ presuming it and $\tau$ are co-axial (Bott, 1959; Wallace, 1951). This same information is obtained from earthquake focal mechanisms, although the ambiguity of the fault and auxiliary planes must be considered. Michael (1984) further made the problem tractable by 1) by assuming that $|\tau_k|$ takes the same value on each fault in the ensemble; 2) by normalizing $|\tau_k| = 1$, and; 3) by considering only the deviatoric part of the stress tensor $\hat{\sigma}_{ij}$. This last assumption necessarily provides an additional useful constraint because $\text{tr}(\hat{\sigma}_{ij}) \equiv 0$ allowing the deviatoric stress tensor to be represented by only five components expressed in a vector $\boldsymbol{t} = [\hat{\sigma}_{11}\ \hat{\sigma}_{12}\ \hat{\sigma}_{13}\ \hat{\sigma}_{22}\ \hat{\sigma}_{23}]^T$. Taken together with $\boldsymbol{a}_k$ being a 3 × 5 matrix comprised of 15 $\boldsymbol{n}_k$ dependent elements Michael (1984) stacked these for the ensemble of *m* faults

$$\begin{bmatrix} s_1 \\ s_2 \\ \vdots \\ s_m \end{bmatrix} = \boldsymbol{S} = \begin{bmatrix} a_1 \\ a_2 \\ \vdots \\ a_m \end{bmatrix} \boldsymbol{t} = \boldsymbol{A}\boldsymbol{t} \tag{7}$$

that may then be solved inversely to find $\boldsymbol{t}$. The resulting normalized deviatoric stress components in $\boldsymbol{t}$ cannot reveal the actual stress magnitudes but do yield both the stress tensor's orientation and R value.

*Table A1. Table of symbols*

| | |
|---|---|
| $\sigma_{i,j}$ | Cauchy stress tensor in Einstein notation |
| $\sigma_1, \sigma_2, \sigma_3$ | Principal components of a stress tensor |
| $S_h, S_V, S_H$ | Andersonian minimum horizontal stress, vertical stress, maximum horizontal stress, in MPa |
| $P_p$ | Formation rock pore pressure, in MPa |
| $P_w$ | Drilling mud pressure, in MPa |
| $\phi$ | Orientation of $S_H$ from North in degrees, stress orientation |
| $\rho$ | Rock density, in kg/m$^3$ |
| R | Shape-ratio of stress |
| $\psi$ | Friction angle of rock, in degrees |
| $\mu$ | Coefficient of friction |
| $C_o$ | Unconfined Compressive Strength, in MPa |
| **T** | Traction vector describe the forces on an arbitrary plane caused by the stress field |
| $\tau_i$ | Shear projection of the traction vector on an arbitrary plane in Einstein notation |
| **t** | Vector consists of 5 elements describing the deviatoric part of a normalized stress tensor |
| **A** | Kernel matrix describing linear relationship between in-situ stress and fault's slipping direction. |
| **s** | Unit vector describing slipping direction of fault |
| **N** | Unit vector normal to the fault |
| $\sigma_i$ | Normal pressure vector in Einstein notation |

## 6 Supplementary materials

Supplementary material is provided electronically including1. a spreadsheet with a summary of analyzed image logs and transient well testing results; 2. grid files for the maps of $\phi$, $S_V$, $S_h$, $P_p$ shown in Fig. 5a, 6d, 8a, 9a; 3. files used for Matlab$^{TM}$ program threeD_stress_duvernay.

## 7 Acknowledgment

The authors would like to thank the anonymous reviewers for their careful reading and constructive feedbacks. The authors also thank the Alberta Energy Regulator for allowing the development and publication of this work. D. R. Schmitt's early contribution were supported by Helmholtz-Alberta Initiative and NSERC Discovery Grant.